\documentclass[%
 aip,
 amsmath,amssymb,
 reprint,%
]{revtex4-2}

\usepackage{graphicx}
\usepackage{dcolumn}
\usepackage{bm}

\usepackage[utf8]{inputenc}
\usepackage[T1]{fontenc}
\usepackage{mathptmx}
\usepackage{etoolbox}
\usepackage{caption}
\usepackage{subcaption}

\makeatletter
\def\@email#1#2{%
 \endgroup
 \patchcmd{\titleblock@produce}
  {\frontmatter@RRAPformat}
  {\frontmatter@RRAPformat{\produce@RRAP{#1\href{mailto:#2}{#2}}}\frontmatter@RRAPformat}
  {}{}
}%

\makeatother
\begin{document}

\preprint{AIP/123-QED}

\author{K. Joeris}
\email[The author to whom correspondence may be addressed: ]{kolja.joeris@uni-due.de}

\affiliation{University of Duisburg-Essen, Faculty of Physics, Lotharstr. 1, 47057 Duisburg, Germany}
\author{L. Schönau}
\affiliation{University of Duisburg-Essen, Faculty of Physics, Lotharstr. 1, 47057 Duisburg, Germany}
\author{M. Keulen}
\affiliation{University of Duisburg-Essen, Faculty of Physics, Lotharstr. 1, 47057 Duisburg, Germany}
\author{J.E. Kollmer}
\affiliation{University of Duisburg-Essen, Faculty of Physics, Lotharstr. 1, 47057 Duisburg, Germany}

\date{\today}

\title[Ultra Low Velocity Ejecta]{Ultra Low Velocity Ejecta Generated by Slow Impacts on Rubble Pile Asteroids }
\date{May 2024}

\begin{abstract}
    We examine ejecta generated by ultra low velocity impacts under asteroid conditions. In an environment of precisely controlled milligravity and under vacuum, impacts with velocities in the range of centimeters/second are performed with irregularly shaped impactors onto granular beds. The resulting ejecta velocities are compared to existing literature values and extend the observed systematic trends towards lower impact energies, broadening the parameter range. Simulations are performed to reason the systematics and the absence thereof for measurements performed at earth gravity. We find, that the cutoff induced by gravity dependent minimal observable velocities plays a crucial role in the values obtained for mean ejecta velocities.
\end{abstract}

\maketitle

\begin{figure}[ht]
     \centering
      \begin{subfigure}[t]{0.45\linewidth}
         \centering
         \includegraphics[width=\linewidth]{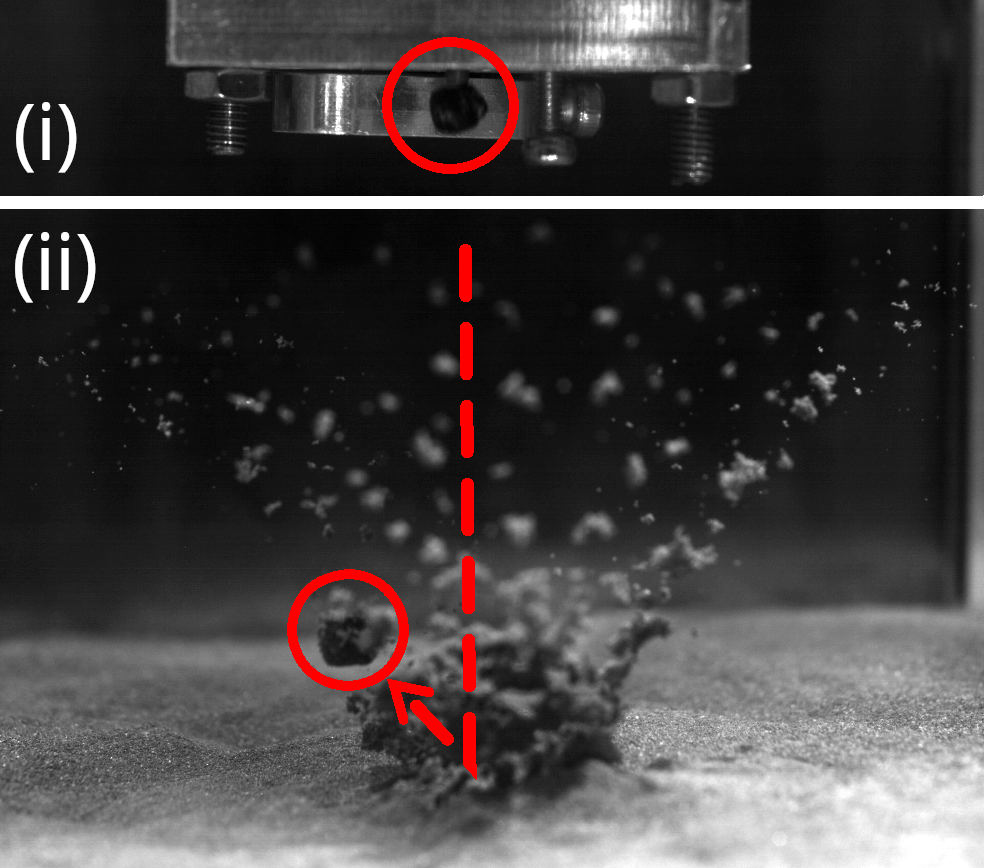}
         \caption{Example impact with two combined images from different time steps, divided by a vertical white line. \textbf{i}: Shortly after launch,   \textbf{ii}: After impact. With impactor in red circle and trajectory as dashed line.}
         \label{fig:yimpact}
     \end{subfigure}
     \hfill
     \begin{subfigure}[t]{0.45\linewidth}
         \centering
         \includegraphics[width=\textwidth]{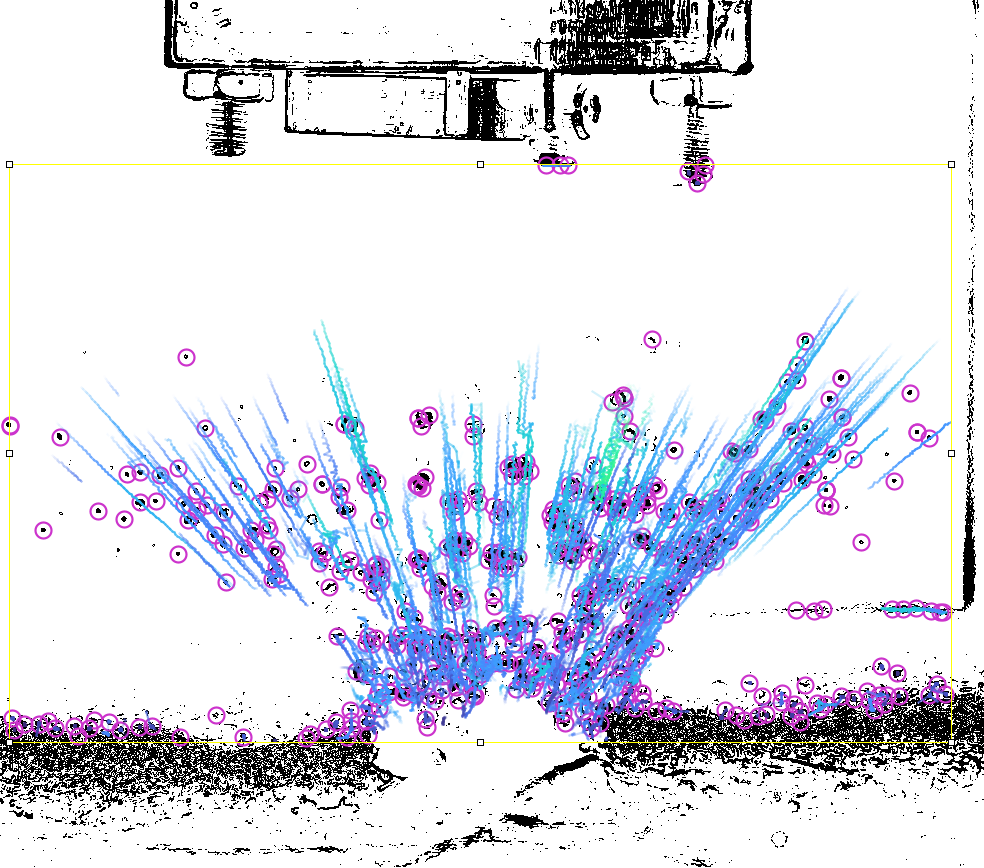}
         \caption{Example impact, tracked with conventional methods using the ImageJ distribution FIJI and the plugin TrackMate. Circles in violet denote identified particles and blue lines are associated tracks.}
         \label{fig:ttracks}
     \end{subfigure}
      \hfill
     \begin{subfigure}[t]{0.9\linewidth}
    \centering
    \includegraphics[width=\linewidth]{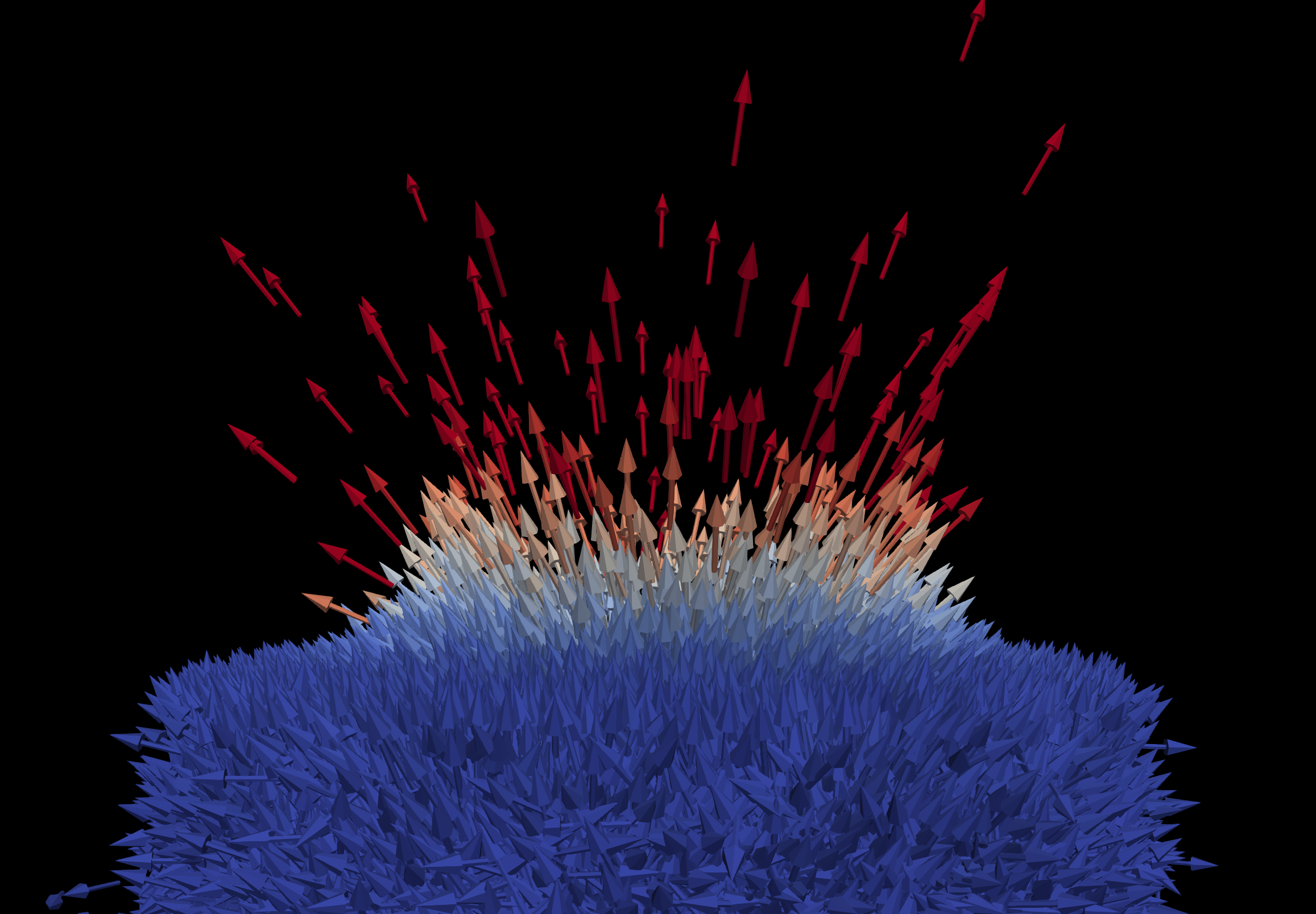}
    \caption{Simulated impact with particles denoted with arrows, color coded according to velocity.}
    \label{fig:rendersplash}
    \end{subfigure}

        \caption{Example ejecta plumes. Views from the main camera (a,b), one with two different time points and one with tracks and simulations (c).}
        \label{fig:maincam}
\end{figure}

\section{Introduction}

Impact processes on celestial bodies have been studied experimentally and numerically extensively, but came into focus even more in the light of recent asteroid missions. Those missions delivered detailed images from asteroid surfaces, fueling research to understand formation and evolution of those objects. More than that, some even performed touchdown maneuvers \citep{yano2006touchdown}, sample returns \citep{lauretta2021osiris} and impact experiments \citep{arakawa2020artificial}. While those missions are extremely insightful concerning for example ejecta generation \citep{li2023ejecta} and momentum transfer \citep{cheng2023momentum}, they come with an exceptionally high prize tag and low repetition rate. As a consequence, the main source for experimental data are ground based experiments \citep{Bogdan,colwell2008ejecta,brisset_regolith_2020,sunday2016novel}, backed by simulations \citep{murdoch2015asteroid}. Our experiments are focused on impacts into weakly bound granular surfaces, with the setup's aim to recreate the dynamics found on rubble pile asteroids \citep{walsh2018rubble, Fujiwara,miyamoto2007regolith}. With a surface gravity in the range of few mm/s$^2$ or lower, the differences to granular dynamics on earth cannot be neglected. First, escape velocities will be small on those bodies \citep{jiang2020motion}. As a result, ejecta generated by (fast) impacts will only re-impact the asteroid, if their ejection velocity is small as well, ranging in the cm/s regime. This filtering effect determines our focus on low velocity interactions. For those slow impacts under asteroid gravity, the force balance changes from what we observe in an environment with earth gravity \citep{scheeres_scaling_2010}. With low gravity and low kinetic energy, the cohesive forces gain importance even for larger particle sizes. With the the increased contribution of cohesive forces, we cannot expect that scaling models that have been derived for higher energies will be readily generalizable to work on extremely weak gravitating rubble pile asteroids. Our analysis thus aims to extend the understanding of granular dynamics into this low energy, high cohesion regime. Granular matter in low gravity is extremely prone to experimental disturbances by even small forces and vibration, thus difficult to explore. While we already studied impactor rebound and regolith surface elasticity in an experiment that created an asteroid environment \citep{joeris2022influence}, we now focus on the ejecta velocity distribution. Both topics are interconnected however, as the ejecta's velocity determines its trajectory and possible subsequent impact events.
\\
To investigate the ejecta velocities, we performed low gravity impact experiments as well as DEM simulations that include inter particle cohesion. We then compare the ejecta velocities obtained in experiment and simulation to literature values \citep{brisset_regolith_2020}. The experiments were performed at the ZARM (Zentrum für angewandte Raumfahrttechnik und Mikrogravitation) Bremen drop tower. Defined asteroid gravity was created by a linear stage performing a constant acceleration to the sample chamber while the stage with the attached chamber were in free fall.  This "linear stage in drop capsule" approach enables us to realize an environment of controlled asteroid gravity $a_a=2 \cdot 10^{-3}\,$m/s$^2$. With a vacuum chamber surrounding the experiment volume, the asteroid simulation is completed. Inside of this volume, a granular bed - mimicking the asteroids regolith covered surface - is located. During an experiment an impactor is launched onto the granular bed, while being observed by multiple cameras. See Fig. \ref{fig:maincam} for (a) the main cameras point of view, (b) particle tracks and (c) numerical simulation. For further details on the experiment and details on the DEM simulation, see the Methods section \ref{sec:methods} and the dedicated publication \citep{joeris2024controlledpartialgravityplatform}.
\begin{figure}
	\centering
	\includegraphics[width=0.45\textwidth]{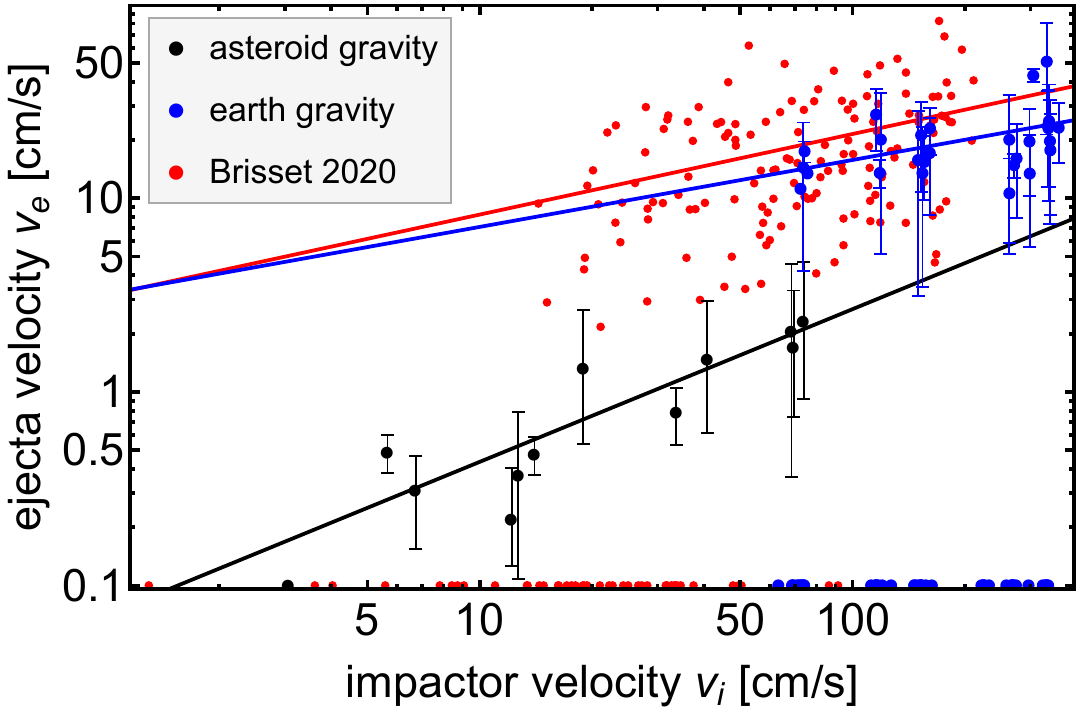}
 \caption{All data points for ejecta velocities from literature \citep{brisset_regolith_2020} partial gravity experiments (red), our asteroid gravity $a_a=2 \cdot 10^{-3}\,$m/s$^2$ experiments (black) and our earth gravity experiments (blue).}
 \label{fig:all}
\end{figure}

\section{Results}
As an Overview, Fig. \ref{fig:all} presents all data of low velocity ejecta generation available to us. Shown here are the ejecta velocities $v_e$ for different impactor velocities while neglecting all other parameters like bed particle size and impactor size. As this is a logarithmically scaled plot with respect to ejecta velocities, points with a value of $v_e=0$ cannot be consistently displayed. Just for this plot, impacts which create no measurable ejecta are not omitted but assigned a velocity of $v_e=0.1cm/s$, making them clearly distinguishable from data points with visible ejecta. As a source for literature values of comparable experiments, we rely on values published by Brisset et al \citep{brisset_regolith_2020} shown in red. The Brisset data points are compiled from different experiments, including parabolic flights, space shuttle flights and a unique laboratory drop tower \citep{brisset_regolith_2020,brisset_regolith_2018,colwell2008ejecta,colwell2003low}.

\begin{figure}
    \centering
    \includegraphics[width=0.9\linewidth]{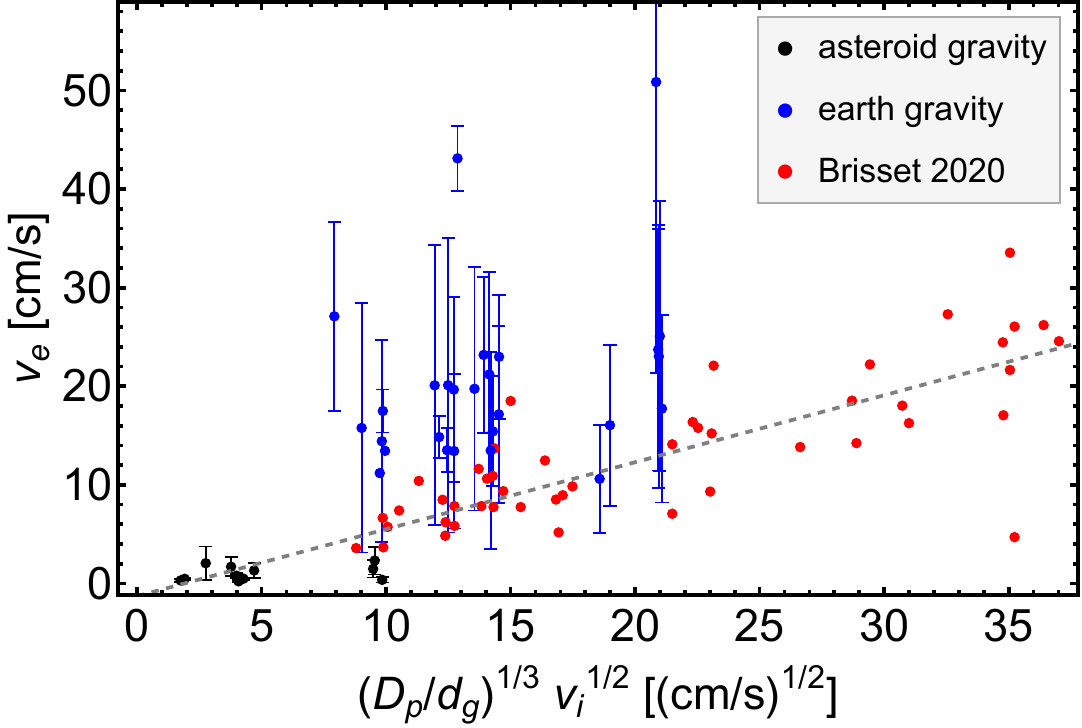}
    \caption{Experimental ejecta velocities $v_e$ with scaled abscissa. Blue Points: Our data in earth gravity. Black points: Our data in asteroid gravity. Red Points: Data collected by Brisset et al \citep{brisset_regolith_2020}. Dashed gray line: Linear fit to combined asteroid gravity data from both experimental sources.}
    \label{fig:scaled-allp}\label{fig:scaled-mug}
\end{figure}

For our data points, we have included both microgravity based experiments recreating asteroid environments (we forth call milligravity data) and earth gravity control experiments. Our data points are shown in blue for impacts under earth gravity and black for asteroid gravity in Fig. \ref{fig:all}. While it seems reasonable to fit a linear function to our milligravity data, all other points spread wildly in this graph. There is no obvious systematic connection between the literature values and our data points. However, the picture changes completely when introducing a scaling similar to Brisset et al. \citep{brisset_regolith_2020}, by taking into account the ratio of impactor size $D_p$ and the bed particle size $d_g$ and the square root of the impactor's velocity. What follows from that for all of our data points combined with those from \cite{brisset_regolith_2020} is shown in Fig. \ref{fig:scaled-allp}. The values taken at earth gravity, plotted in blue, separate more clearly from our (black) and literature (red) microgravity values. On the other hand, when excluding earth gravity values,  a much clearer trend emerges, as shown in Fig. \ref{fig:scaled-mug}. Our data for ultra low impactor velocities extend the point cloud of ejecta velocities towards lower energies. A linear fit matches the combined set of data points closely, verifying Brisset's semi-empirical law for an extended parameter range.

\begin{figure}
    \centering
    \includegraphics[width=0.9\linewidth]{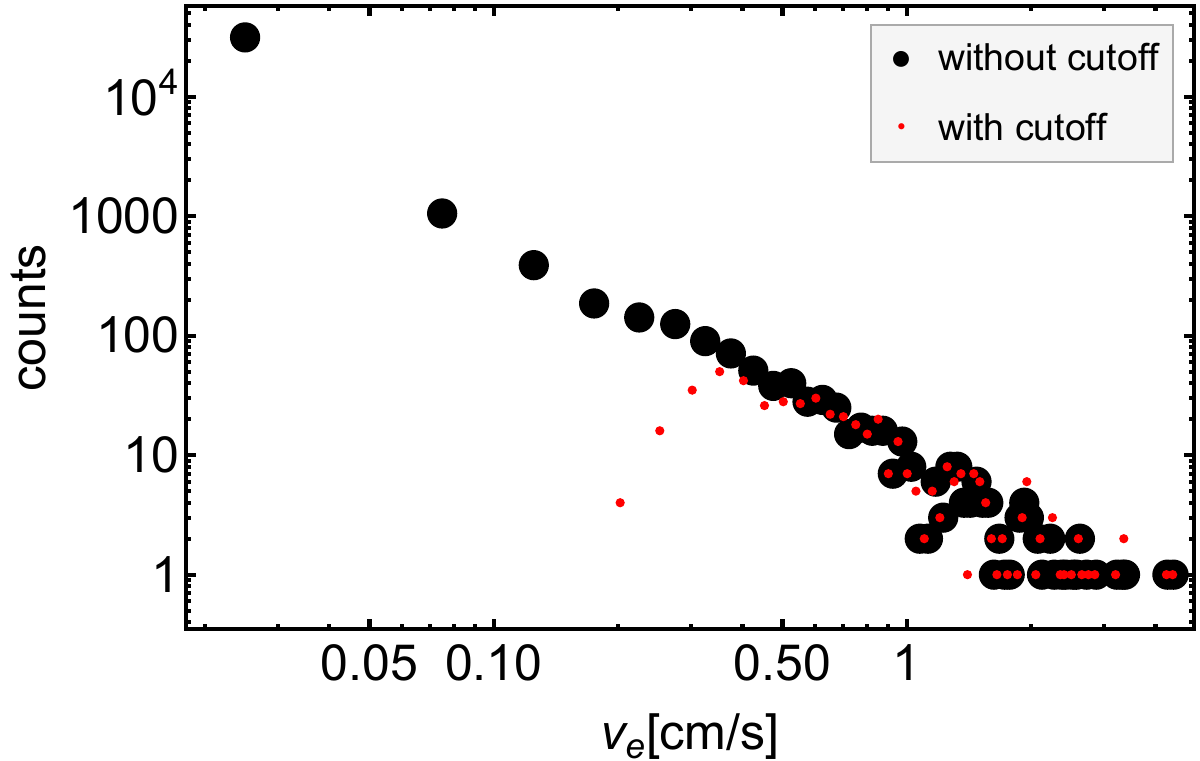}
    \caption{Velocity distribution of ejecta from Simulation, binning with $0.05\,$m/s. Black: All velocities including granular bed. Red: Ejecta velocities with standard cutoff enforced.}
    \label{fig:sim-distro}
\end{figure}

It is obvious from Fig. \ref{fig:scaled-mug}, that the data points we obtained from microgravity experiments align approximately with the literature values while extending it towards lower impactor energies. A change in the ejecta velocity behavior might be expected for low impactor energies under asteroid gravity, because cohesion becomes increasingly important for granular dynamics as shown in \citep{joeris2022influence}. Still, we see no indisputable sign that we have entered this regime just by looking at ejecta velocities. This scaling on the other hand breaks when taking into account our data obtained under earth gravity (blue). No reasonable linear fit can be conducted here, with a low R-Squared of $0.0924$ for the best fit. To investigate the reason for this discrepancy between gravity levels we use the DEM simulation. In contrast to the experiment, the DEM simulation gives us access to the full ejecta velocity distribution, see Fig. \ref{fig:sim-distro}. The bold black dots in Fig. \ref{fig:sim-distro} denote the velocity distribution without cutoff. The black dots following a line in this double logarithmic plot indicates that the velocity distribution follows a power law.  The small red dots in Fig. \ref{fig:sim-distro} are the velocities of a sub-selection of particles, notably those particles that travel above the original bed height. We chose this selection to create a velocity distribution only from particles that would be observable as ejecta in an experiment, in contrast to particles just wiggling around in the bulk regolith below the surface.

\begin{figure}
	\centering
	\includegraphics[width=0.9\linewidth]{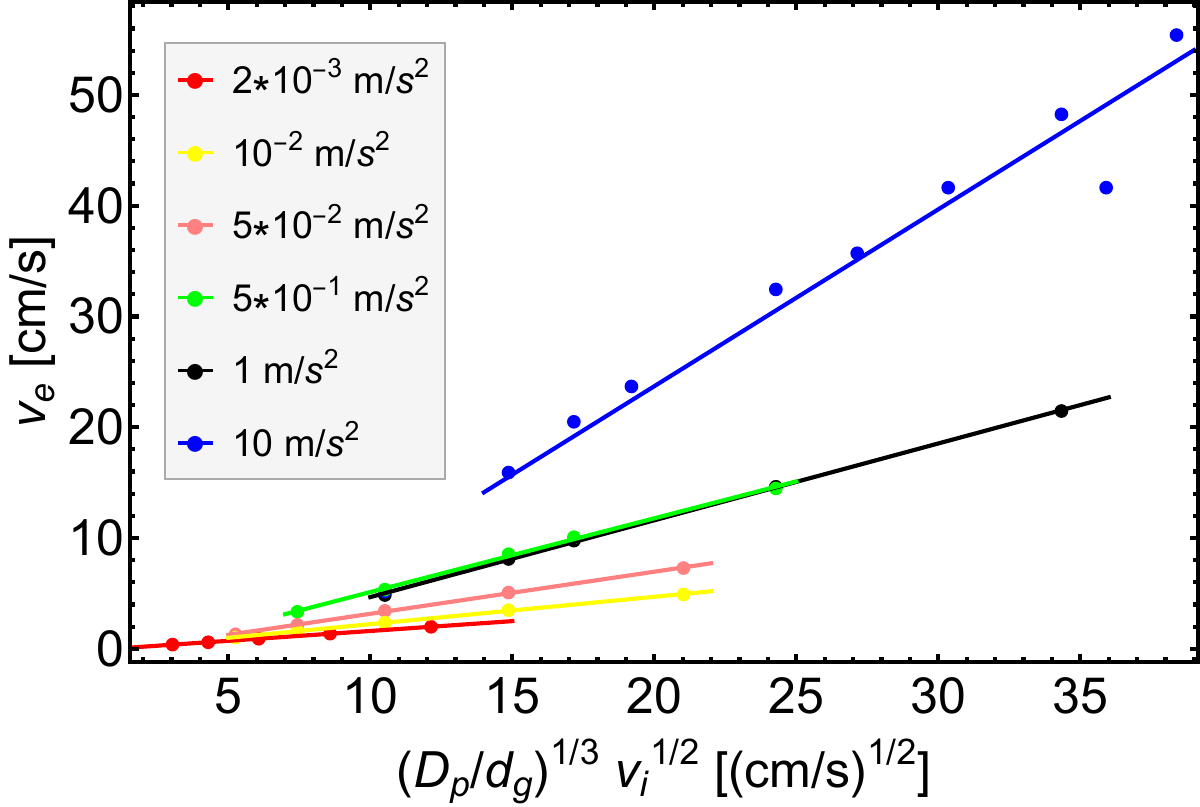}
 \caption{Ejecta velocities taken from numerical simulations $v_e$ with scaled abscissa. Six different accelerations from $2\cdot 10^{-3}\,$m/s$^2$ to $10\,$m/s$^2$ with linear models fitted.}
 \label{fig:sim-all}
\end{figure}


\begin{figure}
    \centering
    \includegraphics[width=0.9\linewidth]{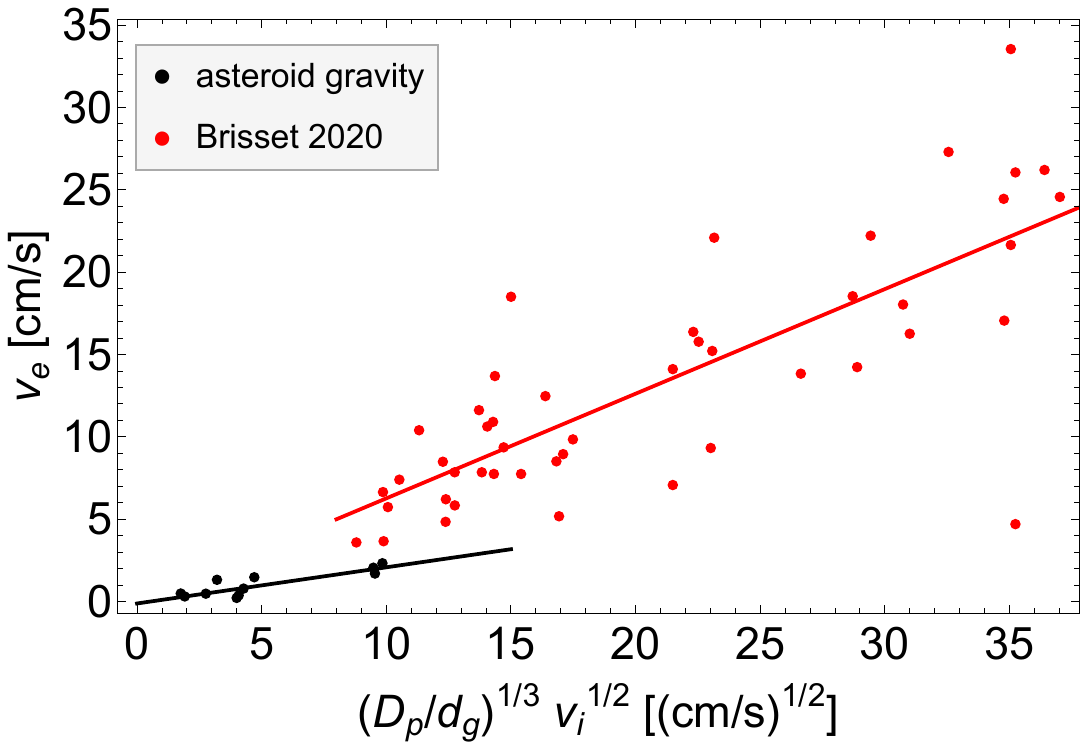}
    \caption{Low gravity experimental data compared in detail. Red: Brisset 2020. Black: Asteroid gravity $2\cdot10^{-3}\,$m/s$^2$}
    \label{fig:lowg}
\end{figure}

The simulations result with the ejecta only cutoff applied is shown in Fig. \ref{fig:sim-all}, displaying the ejecta velocities with the abscissa scaled in the same style as in Fig. \ref{fig:scaled-mug}. Six different gravity regimes are covered from asteroid gravity  $a_a=2 \cdot 10^{-3}\,$m/s$^2$ to earth gravity $a_e=10^4\,$m/s$^2$. The distributions are evaluated at a time point, where the ejecta plume has fully formed. This time is gravity dependent and determined from the initial velocity a particle would need under the given environmental acceleration to reach a height of its own diameter, i.e. be observably displaced. The observation time $t_o=\sqrt{d/a}$ with diameter $d$ and acceleration $a$ is determined from that by dividing the particle radius by this velocity. As seen in Fig. \ref{fig:sim-all}, ejecta velocities in the reduced gravity regimes exhibit a linear behavior for each gravity regime. But like in the experiment, the scaling is not universal for all gravity regimes, the linear fits do not match. Additional parameters might be necessary to collapse the data onto each others. Under earth gravity, no observable ejecta are produced for impactor velocities of $120\,$ cm/s or lower. Fits to the simulated data follow a clear trend with respect to the ambient gravity. As displayed in Fig. \ref{fig:slope}, the slope increases with gravity. This in turn gives reason to compare slopes of the two subsets of reduced gravity data, as shown in \ref{fig:lowg}. The individual fit to our data set yields a slope of $m_a=0.22$, while the fit to the literature values yields a $m_l=0.635$. Compared to numerical values those values can be attributed to simulated environmental accelerations. The value for $m_a$ lies between the slopes for $0.002\,$m/s$^2$ and $0.01\,$m/s$^2$ with $m_{0.002}=0.178$ and $m_{0.01}=0.247$, while $m_l$ lies close $m_{0.5}=0.666$ and $m_{1}=0.694$, within the uncertainties of those fit parameters. The vertical range of the red red area in Fig. \ref{fig:slope} denotes the gravity range of $10^{-4}\,$g to $10^{-2}\,$g \citep{brisset_regolith_2018}.

\section{Discussion}
\begin{figure}
	\centering
	\includegraphics[width=0.9\linewidth]{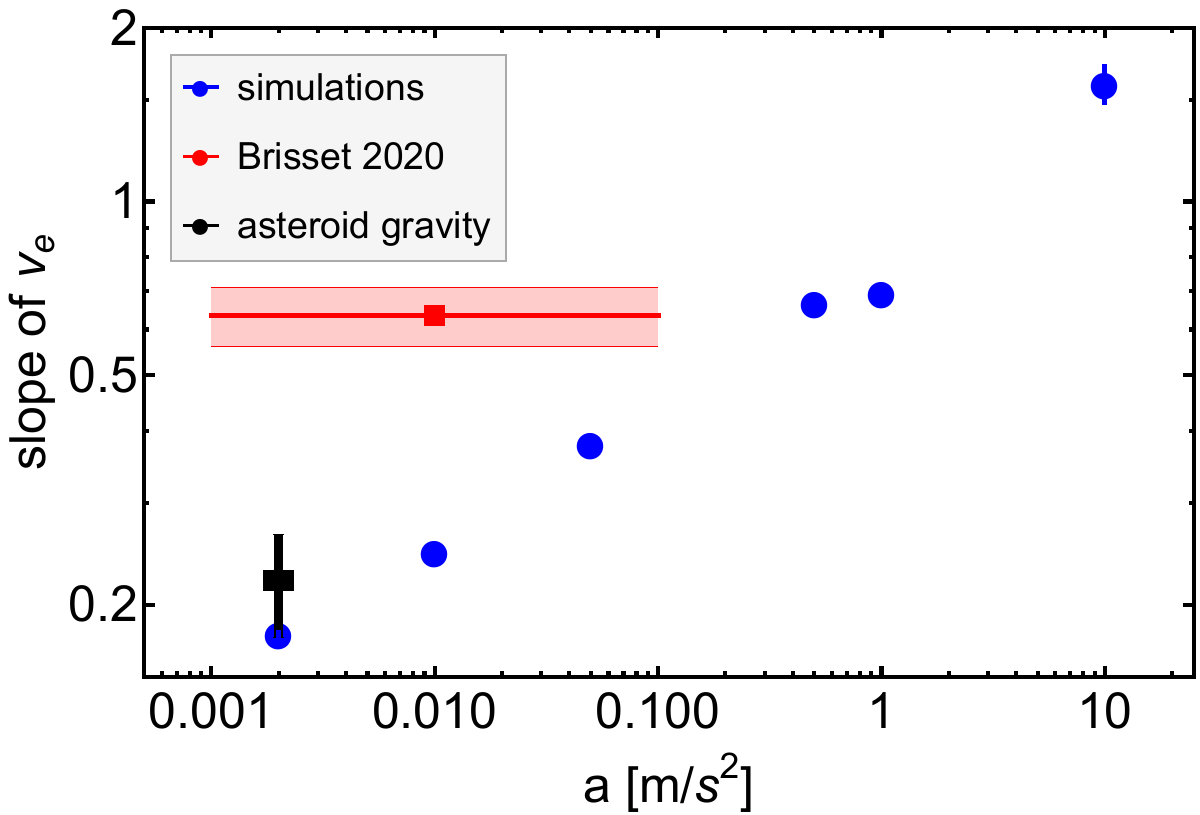}
 \caption{Slopes of linear fits from \ref{fig:sim-all}. Blue: Simulations. Red: Data from Brisset 2020 \citep{brisset_regolith_2020}. Black: Data from our experiment. Uncertainties from standard error of fit parameters. The vertical scale of the red area denotes the range of accelerations given in \citet{brisset_regolith_2018}. }
 \label{fig:slope}
\end{figure}

Our experiment extents the previously known velocity distribution of ejecta to even lower impact velocity. This extension allows to test the ejecta velocity scaling found by Brisset et al. We show, that the linear trend in Fig. \ref{fig:scaled-mug} for ejecta velocities extends into the regime of extremely slow impacts at low partial gravity. 
This trend is not observed at earth gravity.
While the experimental earth gravity data comes with a higher uncertainty, it is set apart from the low gravity results in a similar way in our simulations. We now argue, that one reason for why the Earth gravity data does not fit into this scaling is due to the fact that gravity sets a lower limit for observable ejecta velocities, thus in turn shifting observed mean velocities towards higher values. Since this cutoff is also applied to the simulation data, the observed similarity supports this hypothesis.
Comparing the scaled experimental data from our experiments to the experiments featured in \citet{brisset_regolith_2020} (Fig.\ref{fig:lowg}), we notice a difference if we fit the linear model to the data sets individually. Trying to explain this, we show the slopes of those models in Fig. \ref{fig:slope} together with the slopes obtained from simulation data at different gravity levels. We also indicate the range of gravities under which the experiments were performed. Our data agrees with the simulation data withing margin of error. For the literature data, our simulation data does not agree as well but would suggest either  a higher gravity level, a different observed velocity cutoff or observation time.

\section{Methods} \label{sec:methods}
\subsection{Experiment Methods}\label{sec:exmeth}
This section gives a short overview of the experiment. For details and performance evaluation of the experimental setup and carrier platform see \citet{joeris2024controlledpartialgravityplatform} \cite{Joeris, joeris2022influence}. The core idea is the accurate recreation of asteroid conditions in a controlled environment. To fulfill those requirements, we conduct our experiment at the ZARM drop tower in Bremen. The overall structure of the experiment is as follows: In the evacuated drop tower, a pressurized capsule is launched vertically by a catapult. The capsule is then in free flight and thus experiencing zero gravity. The drop tower guarantees a very clean microgravity with residual accelerations of $10^{-6}\,$m/s$^2$ for $9.2\,$s \citep{von2006new}. Inside the free falling drop tower capsule, our vacuum chamber is attached to a linear stage. This linear stage drives with a constant acceleration of $a_a=2 \cdot 10^{-3}\,$m/s$^2$, creating a controlled asteroid gravity environment. Inside of the experiment vacuum chamber, a launcher is located, which is able to haul impactors onto the asteroid simulant surface at the bottom of the chamber. Each impactor is a irregularly shaped basaltic particle with a diameter of $\approx 3\,$mm. The whole scene is observed with cameras from three different angles. The main camera observes the front of the chamber, perpendicular to the impactor's line of flight and slightly angled to the granular bed's surface. For a view from the main camera, see Fig. \ref{fig:maincam}. The left panel, Fig. \ref{fig:yimpact} shows an example impact with  some highlighted main components. It is a combination of two images from different time points, split by a horizontal white line. The top part, marked \textbf{i}, shows the launcher assembly right after the launch. The impactor, marked with a red circle, is being pushed out of its compartment by a rod moved by a solenoid. While the impactor is retained until shortly before its launch, in the depicted situation the hatch used to retain the impactor is moved out of its way and positioned behind the solenoid rod and the impactor, itself marked with a white outline. The lower part of the image shows a later time frame, marked with a white \textbf{ii}. Clearly visible is the rebounding impactor, again marked with a red circle. Its trajectory is marked with a dashed red line. Around the site of the impact, a plume of ejecta can be observed. To make the ejecta plume's development more clear, Fig. \ref{fig:ttracks} is included. Here, we show an automated tracking preformed on a binarized image sequence. The trajectory of each particle is shown in blue, with the length hinting on its velocity and the particle itself marked in red. Limitations of this method are already apparent: Many particles are wrongly detected and deviations from the actual smooth trajectory yield a systematic overestimation of velocities for this particle size. Other methods for ejecta characterization are detailed below.

\subsubsection{Particle Velocity Determination \label{sec:track}}
The impactor's velocity is measured manually in all cases, by tracking it using the main camera. The ejecta velocities are measured using the main camera located at the chamber's front, with optical axis near perpendicular to the impactor's flight direction and the simulant surfaces normal vector. From the image data obtained with this setup, the velocities are extracted using three different methods. \\ \\
The simplest one is the direct tracking method. Here, particles are located individually using footage from the main camera and identified individually and manually over as many frames as possible. This method only works for targets composed of larger grains and ejecta populations which consist of only a few grains. For those cases of course it becomes most reliable and accurate and can provide a benchmark for the accuracy of other velocity field measurement methods. The error of this method is determined by spatial and temporal resolution. With an increasing number this method becomes less feasible. Still, with ejecta numbers below $30$, a different variant of this method can be used, but the number of observed frames has to be decreased. Here, the velocity is determined from only two frames and averaged over all particles. As a third variant, useful for small particles, what we call the radius method is employed. Here, a half circle is positioned around the impact point. Then the time is measured from impact until the ejecta passes the half circle. For each impact, this was done using each three radii. \\ \\
Automated tracking does not work well for our irregular, low contrast particles. An example image with many incorrectly identified particles is shown in Fig. \ref{fig:ttracks} . But while individual particles cannot be resolved reliably, the flow field yields information. So, in cases with strong ejecta production and small bed particles, particle imaging velocimetry (PIV) is used. PIV is a technique widely used in fluid dynamics, usually employed to examine flows of fluids or gasses. Images of tracer particles inserted into the examined flowing medium are taken at two time points. The image is then partitioned into interrogation regions and both images cross-correlated, to find a two dimensional flow vector. Applying the technique to our data is straight-forward: An image pair is composed of two consecutive images from our video data. We dot not examine a flowing medium, but are interested in the particles themselves. Specifically, we use the PIV plugin distributed with the open source image processing package \textbf{Fiji}. The time interval in which the splash occurs is isolated and the frame rate reduced to leave $20$ frames. PIV is performed for each two consecutive images. The interrogation window is chosen to $1/8$ of the original image dimension, with the search window double its size and the vector spacing half its size.  The next interrogation window sizes are reduced by a factor of $1/2$ for each iteration. For further investigation the third iteration is used.

\subsection{Simulation Methods \label{sec:simmth}}
The physical system was additionally recreated in a numerical simulation. We used the LIGGGHTS software package \cite{kloss2012models}, an open source project for discrete-element methods. Grains are modeled as spherical particles. In the simulation, the experiment was recreated. To achieve this, a set of $32891$ spherical particles with diameters of $0.9\,$mm radius is dropped into a cylindrical container with a diameter of $10\,$cm and a depth of $2\,$cm. Settling of the granular bed is performed a gravity of $10\,$m/s$^2$, close to earth gravity, to better resemble the experiment. Before insertion of the impactor, the gravity is reduced to the targeted gravity of $a_a=2 \cdot 10^{-3}\,$m/s$^2$ or kept at $10\,$m/$s^2$ for earth gravity control measurements. After settling, an impactor is inserted above the granular bed's surface with a given initial velocity with only one non vanishing component parallel to the gravity vector, essentially launching it in the direction of the granular bed.
For the contact model a dissipative Hertzian model was chosen (\texttt{gran model hertz} in LIGGGHTS)  with a linearized variant of Johnson-Kendall-Roberts \cite{JKR} cohesion (\texttt{SKJR2}). Further simulation parameters chosen are the following: Particle density was set to $2.1\,$g/cm$^3$, which is in the range of the bulk density of silicates like soda lime glass or basalt \cite{Steinpilz20,le1991iugs}, while allowing some additional porosity. The Young's modulus is set to $10^9\,$Pa, which is about one to two orders of magnitude lower than that of real basalt \citep{vennari2021deformation} to allow for smaller simulation time steps and a shorter simulation time. The Poisson ratio is set to $0.2$, the coefficient of friction to $0.2$, the coefficient of restitution to $0.5$ and the cohesion energy density to $21600\,$ergs/cm$^3$. 
 \section{Author declarations}
 \subsection{Conflict of Interests}
 The authors have no conflicts to disclose.
\subsection{Acknowledgements}
This work was supported by the DLR Space Administration with funds provided by the Federal Ministry for Economic Affairs and Climate Action (BMWK) based on a decision of the German Federal Parliament under grant number 50WM1943, 50WK2270C and 50WM2243.
\subsection{Author Contributions}
  \textbf{Kolja Joeris:} Investigation (lead); Resources (lead); Software (lead); Writing - original draft (lead); Visualization (lead); \textbf{Laurent Schönau:} Investigation (supporting); ; Writing - original draft (supporting); Visualization (supporting). \textbf{Matthias Keulen:} Investigation (supporting); Writing - original draft (supporting); Visualization (supporting). \textbf{Jonathan E. Kollmer:} Conceptualization (lead); Supervision (lead) Project Administration (lead); Funding Acquisition (lead); Writing - original draft (supporting).

\bibliography{bib}
\end{document}